# The End-of-Day Reversal After 2019: A Pre-Registered Test on S&P 500 Stocks Excluding Heavily Traded Names

Alex Vidovich[a,*], Zivana Zerjal[b]

[a] *Independent researcher · alexvidovich.com*
[b] *Independent researcher · zivanazerjal.com*

* *Corresponding author.*
*E-mail addresses:* support@alexvidovich.com *(A. Vidovich)*, support@zivanazerjal.com *(Z. Zerjal)*

ABSTRACT

Baltussen, Da and Soebhag (2025) document an end-of-day reversal in U.S. stocks through December 2019: intraday losers outperform intraday winners over the last half hour of trading. We test it on later data, on point-in-time S&P 500 members excluding about 80 heavily traded stocks we had examined in earlier, unpublished work. Under a private pre-registration, the 2016–2019 data had to reproduce the published sign before any statistic linking the sort to the outcome was computed on later primary-sample data. Measured to the last trade before the closing auction, the dollar-volume-weighted reversal is 3.13 bps per day (t = 7.5) in 2016–2019 and 3.05 bps (t = 3.0) in 2023–2026, with a one-sided 95% bound on the decline of 61%. The result depends on the window: the reversal is indistinguishable from zero in 2020–2022, 2023 carries much of the later estimate, and from 2020 onward it is 1.69 bps (t = 2.3), with a bound of 91%. A proxy sample, added after the run, gives 2.81 and 3.86 bps. On the full index the reversal falls from 3.40 to 0.74 bps (t of the change −1.83, short of the registered bar of −3); descriptive comparisons concentrate this weakness in the very largest stocks. No pre-registered mechanism test passes.

All figures come from the authors' dated research log and its result files. The design, estimators and decision rules, drafted by an AI model (Section 3.4), were pre-registered before any statistic linking the sort to the outcome was computed on primary-sample data after 2019. The scoring code, the daily point-in-time S&P 500 membership, the leveraged-ETF launch table with an issuer source for every date, public copies of the pre-registration documents, and the code for every post-hoc analysis are released at https://doi.org/10.5281/zenodo.24003093. The price bars are not redistributed, under the vendor's licence; rebuilding the results requires access to Alpaca's consolidated-tape (SIP) market data. The public pre-registration copies redact references to the authors' other projects and the identity of the excluded list, so each has its own SHA-256 hash; a table in the archive lists the original and public hashes. The list of excluded stocks is not released, because it reflects the authors' other, unpublished work, so the primary sample cannot be rebuilt exactly from outside. We reproduced the full-index and proxy-sample results exactly from the released files; no outside party has yet tried.

## 1. Introduction

Baltussen, Da and Soebhag (2025, hereafter BDS) sort U.S. stocks each afternoon on their return from the prior close to 3:00 p.m. and find that, over the last half hour, the day's losers beat the day's winners by 3.78 bps a day in market-capitalization-weighted portfolios (t = 10.69, 1993–2019). They attribute the pattern to attention-driven retail buying of losers and to short sellers who reduce risk before the close. The effect is large for a 30-minute window, and it was documented just before the close of the U.S. market changed in several ways at once: daily index options, single-stock leveraged ETFs and a larger closing auction all arrived or grew after 2019. Each of these could have weakened the pattern, and some could have reversed it.

Out-of-sample attenuation is well documented. McLean and Pontiff (2016) find that returns to published predictors are about 26% lower out of sample and about 58% lower after publication, and large-scale replications find that many anomalies are weaker than first reported once small stocks are given less weight (Hou, Xue and Zhang, 2020), although most survive in a broad international sample (Jensen, Kelly and Pedersen, 2023). BDS's data stop in December 2019, and we are not aware of a published test of the reversal on later data.

We find that the reversal was present in 2023–2026 in the registered sample. On point-in-time S&P 500 members, excluding about 80 large, heavily traded stocks whose recent intraday data we had examined in earlier, unpublished work, the reversal is 3.13 bps a day in 2016–2019 and 3.05 bps a day in 2023–2026 (Table 2), and the one-sided 95% confidence bound on the decline is 61% of the 2016–2019 effect. Because the excluded list is not released, we also report a proxy sample that anyone can rebuild, which drops the 80 largest stocks by prior-month dollar volume each day; it gives a larger estimate for 2023–2026 (3.86 bps, t = 5.3), although it shares most of its stocks with the primary sample and is not independent confirmation. The full index behaves differently: with dollar-volume weights, its reversal is 0.74 bps (t = 0.55) in 2023–2026. Descriptive comparisons, which we did not pre-register, concentrate this weakness in the very largest stocks. Among the 80 largest by dollar volume, the reversal falls from 3.88 bps (t = 5.74) to 0.33 bps (t = 0.18), and the change in the proxy sample minus the change among those 80 is 4.61 bps (t = 2.60). With equal weights, however, the full-index reversal in 2023–2026 is 3.17 bps (t = 4.37), and without the five largest stocks each day it is 2.98 bps (t = 3.11). The weakness is therefore concentrated in a few stocks that dominate dollar-volume weights rather than spread across the index.

The result depends on the window. BDS's sample ends in 2019, and 44% of the later sessions fall in 2020–2022, where the reversal is indistinguishable from zero. We fixed the split into three periods before the run, but the pre-registration gives no reason for it, and it reports 2020–2022 only descriptively. Over all sessions from 2020, the reversal is 1.69 bps (t = 2.27), 46% below 2016–2019, and the one-sided 95% bound on the decline is 91% (Table 3). It also rests on 2023, the strongest year, which differs from 2024–2026 (1.80 bps, t = 1.37; gap t = 2.82), while 2024–2026 cannot be told apart from 2020–2022 (gap t = 1.10), so a statement that the reversal returned in 2023 rests on one year. BDS first circulated their paper in 2024, so the post-publication evidence is limited to 2025 and the first ten months of 2026; we describe the result as out of sample, not as surviving publication.

We also test four candidate explanations for changes at the close: the market and idiosyncratic components of the intraday move, the 2022 introduction of additional same-day index option expiries, and the staggered launch of single-stock leveraged ETFs, whose daily rebalancing trades in the direction of the stock's move. None passes its pre-registered threshold after a Holm correction. The leveraged-ETF test fails by a sample-size rule rather than on its estimate, because only 12 treated stocks lie in our sample and the rule required 20. An exploratory split of that test, chosen after the confirmatory run, points in the predicted direction; we have registered a test of it on later launches.

The analysis was pre-registered, including its estimators, thresholds and the wording of each verdict; an AI model drafted the registration (Section 3.4). A reproduction test came first (on the credibility of published results, see Harvey, 2017): the 2016–2019 data had to show BDS's sign at t ≥ 2 before any statistic linking the sort to the outcome was computed on later primary-sample data. It reproduces the sign only; on the primary sample, BDS's next-morning give-back does not reproduce (Section 4.6). On the full index in 2023–2026, the return from 3:30 p.m. to the next open has the opposite sign, −13.7 bps (t = −2.88); it depends on the dollar weights of a few large stocks (Table 8, descriptive). An adversarial audit of the design, run by an AI agent before the reproduction test, found two defects that would have let the paper overstate a change, and both were fixed before any primary-sample outcome was seen (Appendix B). The universe is point-in-time and includes stocks that were later removed from the index. The design has low power against a decline: at our standard errors, a true 50% decline would produce the "weakened" verdict about 6% of the time and "not proven" about 62% of the time (Table A2).

The reproduction test can serve as a template for out-of-sample work: the in-sample result must be reproduced on the new data source before any out-of-sample statistic is computed. We also release a table of first single-stock leveraged-ETF launch dates for S&P 500 members, each sourced from issuer documents; an aggregator source we first used placed nine launches up to ten months too early. A small data check before the run suggested that the vendor's daily close drifted away from the closing-auction print after 2019, and a full comparison after the run shows that it does not (Section 3.2). The paper does not propose a trading strategy and makes no claim net of costs.

Section 2 describes the anomaly and the forces that could have changed it. Section 3 sets out the data and design, Section 4 the main results, Section 5 the mechanism tests, Section 6 the limitations, and Section 7 concludes.

## 2. The anomaly and what could have changed it

BDS define ROD3 as a stock's return from the prior close to 3:00 p.m., which combines the overnight return, the opening and the middle of the day, and they skip the half hour from 3:00 to 3:30 to avoid bid-ask bounce. Sorting stocks on ROD3, the bottom quintile beats the top quintile over the last half hour, 3:30 to 4:00 p.m. They report that, apart from the 10:00–10:30 interval, the pattern does not appear in earlier half-hours (in our 2016–2019 sample a smaller one appears at 2:00–2:30; Section 4.5), that it is distinct from the market-level intraday momentum of Gao, Han, Li and Zhou (2018), and that it survives controls for option and leveraged-ETF hedging demand. It belongs to a wider literature on intraday return predictability, including the half-hour continuation of Heston, Korajczyk and Sadka (2010) and the overnight–intraday tug of war of Lou, Polk and Skouras (2019). BDS attribute the reversal to positive price pressure on intraday losers, and a reversal of this kind can be read as the return to providing liquidity against order imbalances (Nagel, 2012). Price pressure from order imbalances at the close is older still (Cushing and Madhavan, 2000).

Some flows at the close must trade in the direction of the day's move. A leveraged or inverse ETF with daily leverage multiple L rebalances in proportion to $L(L - 1)$ times the day's return, which is positive for every such fund, so it buys after a rise and sells after a fall (Cheng and Madhavan, 2009), although Ivanov and Lenkey (2018) find that fund flows offset much of this rebalancing. Single-stock leveraged ETFs arrived in the United States in July 2022. Dealers who are short gamma on a stock's options hedge in the same direction, and Barbon, Beckmeyer, Buraschi and Moerke (2022) show that both flows move end-of-day prices and that the effect fades within the next day. At the market level, Baltussen, Da, Lammers and Martens (2021) link intraday momentum in the last half hour to the same hedging flows. Other flows push the opposite way. Dealers who are long gamma hedge against the move, and Dim, Eraker and Vilkov (2024) find that dealers are on average net long the gamma of same-day index options, so these options should dampen intraday moves. The closing auction itself has grown: Bogousslavsky and Muravyev (2023) report that its share of daily volume rose from 3.1% in 2010 to 7.5% in 2018, and its price impact differs across exchanges (Jegadeesh and Wu, 2022), which matters here because our outcome ends before the auction (Section 3.2).

The registered predictions are signs only; the registration states no magnitudes. If continuation flows grew after 2019, the reversal should weaken (Section 4.2). The table maps each force to its test. Section 5 runs the mechanism tests after Section 4 establishes whether there is anything to explain.

*Registered predictions and the tests that address them (reversal terms).*

| force | prediction | test |
|---|---|---|
| index hedging that adds continuation | market-component reversal falls | Table 9, first row |
| dealers long the gamma of index options | market-component reversal rises | Table 9, first row |
| stock-level flows (single-stock leveraged ETFs, option gamma) | idiosyncratic reversal falls | Table 9, second row |
| same-day index options on the weekdays that gained them | two-sided: falls with added continuation, rises with dampening | Table 9, third row |
| single-stock leveraged-ETF launches | reversal falls after launch | Table 9, fourth row; Section 5.2 |
| transitory price pressure | next morning gives back the close | Section 4.6 |

## 3. Data and design

### *3.1 Universe*

The universe is the set of S&P 500 members on each trading day from January 4, 2016, to October 30, 2026. The sample starts in 2016 because the vendor's consolidated-tape history begins then. Membership comes from a community-maintained file of historical index components and changes (fja05680, 2026), pinned to the version of September 7, 2026; changes after August 18, 2026, are taken from the public changes table, and every change was checked against the running set of members. Stocks later removed, acquired or renamed are kept and mapped by company, and one ticker that belonged to two different companies in different years is mapped by company. About 1,970 member-days, or 0.14%, have no usable price data; almost all belong to eight companies acquired in 2016–2019 for which the vendor has no bars, and they are dropped.

The primary sample excludes a fixed set of about 80 large, heavily traded stocks whose recent intraday data we had examined in earlier, unpublished work, matched by company so that earlier tickers are excluded as well. We excluded them so that no stock we had already studied could shape the result. The list reflects which stocks were heavily traded in recent years, so in 2016–2019 it conditions on later trading: it removes stocks because of what they became. The rise in their dollar-volume weight between the periods, reported below, fits this. The proxy sample described below avoids this look-ahead. About 420 stocks remain on a typical day, and all primary tests use them. Because the excluded stocks are among the most heavily traded in the index, they carry 46% of the daily dollar-volume weight in 2016–2019 and 59% in 2023–2026 (Table 1). The full index is a pre-registered robustness check. Because the list is not released, we add a proxy sample, not pre-registered and introduced after the confirmatory run, that drops the 80 largest members by prior-month dollar volume on each day; it uses only public data, and it is reported alongside the primary sample throughout Section 4. The two samples overlap but are not the same: on a typical day roughly 60% of the excluded stocks are among the 80 largest, so roughly 30 of the 80 largest stocks are in the primary sample and roughly 30 excluded stocks are not among the 80 largest. On a typical day the two samples share about 390 stocks, which carry 73–78% of the primary-sample weight and 86–88% of the proxy weight, and their daily estimates correlate 0.91 in 2016–2019 and 0.65 in 2023–2026. The proxy is therefore not independent confirmation.

### *3.2 Prices*

Prices are consolidated-tape (SIP) bars from Alpaca's market-data service: daily bars, both raw and adjusted for splits and dividends, 30-minute bars, and 5-minute bars for the last half hour, which supply the 3:55 p.m. price used in one robustness variant, all placed on the New York clock. The price at 3:00 p.m. is the close of the 2:30–3:00 bar, that is, the last trade before 3:00; the price at 3:30 is the close of the 3:00–3:30 bar; and the end of the outcome window is the close of the 3:30–4:00 bar, the last trade before 4:00. Of the 2,723 sessions from January 4, 2016, to October 30, 2026, we drop 46: the 21 scheduled early closes; six sessions in which fewer than 95% of members have the required bars, which are vendor gaps that a fresh download reproduced; and the 19 sessions of January 2016, for which no prior-month dollar volume exists to weight the stocks. That leaves 977, 747 and 953 sessions in the three periods. We also drop stocks priced under $5 at the prior close, stock-days with an overnight move larger than 40% in absolute value, which flags unadjusted spin-offs and ticker splices and is known at the open, and one company whose vendor data were broken across a share-class change.

The outcome ends at the last trade before 4:00 p.m. rather than at the closing auction. The pre-registration chose this end point because a check of 40 stock-days, made before the reproduction test, found a print on 33 of them and suggested that the vendor's daily close matched the closing-auction print on 12 of 13 days in 2016–2019 but differed by 1–7 bps on 16 of 20 days from 2020 onward. A full comparison after the confirmatory run, on the vendor's auction records (the largest print with condition 6, a rule written before the download), shows that this suggestion was wrong: the vendor's daily close equals the closing-auction print on 99.2% to 100% of stock-days in every year, and the exceptions cluster in 2020. The earlier check took the first print marked as a closing or official-close print, and from about 2020 the feed also carries such prints from other venues at slightly different prices. The registered end point stands, so every registered result is measured before the closing auction. Below, "vendor's daily close" names the registered version, and "auction print" names the rebuilt one. The vendor's daily open is within 1 bp of the opening-auction print on only 66–83% of stock-days. BDS report that their result is robust to skipping the last five minutes and to not using the closing price.

### *3.3 Variables and estimator*

For stock i on day t, let $r_{i,t}^{on}$ be the overnight return from adjusted daily bars, $O_{i,t}$ the vendor's daily open, and $P_{i,t}(h)$ the price at clock time h as defined above. The sorting variable and the outcome are

$$\text{ROD3}_{i,t} = (1 + r^{on}_{i,t}) \frac{P_{i,t}(15{:}00)}{O_{i,t}} - 1, \qquad \text{LH}_{i,t} = \frac{P_{i,t}(16{:}00^{-})}{P_{i,t}(15{:}30)} - 1, \tag{1}$$

where 16:00⁻ denotes the last trade before 4:00 p.m. Because the overnight return is built from the same open, the open cancels in ROD3, which is in effect the return from the prior close to 3:00 p.m.; the open matters only for the overnight screen and the next-morning window of Section 4.6. Rebasing the sort on the prior day's auction print leaves the reversal at 3.13 and 3.08 bps. Each day we rank ROD3 across stocks, scale the rank to $q_{i,t} \in [-0.5, 0.5]$, and estimate by weighted least squares

$$\text{LH}_{i,t} = a_t + b_t\, q_{i,t} + e_{i,t}, \qquad \text{weights } w_{i,t}, \tag{2}$$

with LH in basis points and $w_{i,t}$ equal to the stock's mean daily dollar volume over the prior calendar month. The slope is $b_t = \sum_i w_{i,t}(q_{i,t} - \bar{q}_t)\text{LH}_{i,t} / \sum_i w_{i,t}(q_{i,t} - \bar{q}_t)^2$, where $\bar{q}_t$ is the weighted mean rank. It is the return of a zero-cost portfolio, long stocks ranked above $\bar{q}_t$ and short those below, with positions proportional to $w_{i,t}(q_{i,t} - \bar{q}_t)$ and scaled to unit exposure to the rank. The daily reversal is $R_t = -b_t$, positive when losers beat winners. BDS weight by market capitalization; our dollar-volume weights favour heavily traded stocks even more. Unlike BDS, who winsorize at 1% and 99%, we do not winsorize; Table 7 shows both. For each period P, the reversal is the mean of the daily estimates, $\bar{R}_P = N_P^{-1} \sum_{t \in P} R_t$, with a Newey–West (1987) standard error using five lags, and the change between the early and late periods is

$$\Delta = \bar{R}_L - \bar{R}_E, \qquad \text{se}(\Delta) = \sqrt{\text{se}(\bar{R}_E)^2 + \text{se}(\bar{R}_L)^2}. \tag{3}$$

As a second statistic we report, as BDS do, the return of the bottom ROD3 quintile minus the top quintile, weighted within each quintile by the same dollar-volume weights.

### *3.4 Periods, order of runs and verdicts*

Claude (Anthropic), an AI model, proposed and drafted the design, the pre-registration and the decision rules, and the authors adopted them; an AI agent ran the adversarial audit of Appendix B (see Use of AI tools at the end of the paper). Throughout, "descriptive" marks a result that the registration reports without a bar or that we added after the confirmatory run, often on data we had seen; it tests nothing and enters no verdict, and Appendix D lists each addition. Period E (2016–2019) overlaps BDS's sample, period T (2020–2022) is reported descriptively and enters no verdict, and period L (January 2023 to October 2026) is the registered test period. The pre-registration gives no reason for this split, and the result depends on it (Table 3). The pre-registration fixed the order of runs. The reproduction test was run on E alone and required $\bar{R}_E > 0$ with t ≥ 2. Checks before that run did read later data, including closes from 2020 onward and a data panel (Appendix B), and we had seen recent intraday data on the excluded stocks; no statistic linking the sort to the outcome was computed on later primary-sample data before the confirmatory run. Only after that result was committed was the confirmatory run executed, once, on all periods; the registered script can be executed once, and a committed lock blocks a second execution. The rank transform, the weights, the period boundaries, the Newey–West lag, the 21-day block bootstrap with 999 draws used for the change-point test and the next-morning interval, and the random seed were all fixed in advance, and the code checks the hash of every input before it runs.

The verdicts were written in terms of the slope and are restated here in terms of the reversal (Appendix A gives both). Following the multiple-testing argument of Harvey, Liu and Zhu (2016), the threshold for declaring a change is t = 3. The reversal "weakened" if Δ < 0 with t ≤ −3; it "changed sides" if, in addition, $\bar{R}_L < 0$ with t ≤ −3 and the version ending at the vendor's daily close agrees in sign. The no-weakening verdict holds if Δ ≥ 0, or if $\bar{R}_L > 0$ with t ≥ 2. If both "weakened" and the no-weakening verdict hold, the result is "weakened but still a reversal"; anything else is "not proven". As written, the first clause of the no-weakening verdict could hold with $\bar{R}_L$ close to zero; in our data only the second clause applies, so this does not affect the result. The t = 3 bar comes from an argument about screening many factors and is stricter than a single registered test needs, while the no-weakening verdict and the 61% bound below use lighter bars. We therefore report changes against both the registered bar and a one-sided 5% test (Table 3).

The pre-registration documents and their hashes are in the public archive; they were committed to a private version-control history before the runs they govern, but those commit times cannot be verified from outside and readers must take them on trust. The released copies are redacted where they name earlier work and therefore carry their own hashes; the original hashes prove nothing to an outside reader. The follow-up test described in Section 5.2 is different: it is deposited with the archive, which dates it, and no stock-level result for it has been computed.

## 4. Results

### *4.1 Sample description and reproduction test*

Table 1 describes the three periods. The cross-sectional dispersion of both ROD3 and LH rose after 2019 and stayed higher, which is one reason the daily reversal is estimated less precisely in the later periods.

*Table 1. Sample description (means of daily values).*

| | E, 2016–2019 | T, 2020–2022 | L, 2023–Oct 2026 |
|---|---|---|---|
| Sessions | 977 | 747 | 953 |
| Stocks per day, primary sample (full index) | 420 (503) | 419 (504) | 417 (502) |
| Dollar-volume weight of the excluded stocks | 46% | 56% | 59% |
| Cross-sectional s.d. of ROD3 (%) | 1.42 | 1.82 | 1.74 |
| Cross-sectional s.d. of LH (bps) | 22.8 | 34.3 | 33.3 |
| Weighted mean LH (bps) | −0.82 | −1.37 | −0.45 |

On the 977 sessions of 2016–2019, the reversal is 3.13 bps a day (Newey–West t = 7.52), and it is positive and significant in each of the four years (Table 4). The quintile spread is 2.55 bps a day (t = 7.27). BDS report 3.78 bps (t = 10.69) for market-capitalization-weighted quintiles over 1993–2019, so our quintile spread is about two-thirds of theirs, on a different period, universe, weighting and end point. Size alone does not explain the gap, since BDS's six-factor alpha for their largest firms is 3.41 bps (t = 10.61); our shorter window, our dollar-volume weights and our pre-auction end point are other differences, and we cannot apportion the gap among them. A size check specified in the pre-registration addendum, before the reproduction test, ran the test on 200 within-session shuffles of LH: it passed falsely in 3.0% of draws against a nominal 2.3%; with 200 draws the 95% interval is 1.1% to 6.4%, so this confirms little. The test passed, and a second implementation, written separately from the registered code but by the same AI tool from the same specification, reproduced every figure. It catches coding slips, not shared misreadings of the specification.

### *4.2 The main result*

*Table 2. End-of-day reversal $\bar{R}_P$ by period (bps per day; positive = losers beat winners).*

| | sessions | primary sample | t | proxy sample | t | full index | t |
|---|---|---|---|---|---|---|---|
| E, 2016–2019 | 977 | 3.13 | 7.52 | 2.81 | 6.78 | 3.40 | 6.65 |
| L, 2023–Oct 2026 | 953 | 3.05 | 2.98 | 3.86 | 5.26 | 0.74 | 0.55 |
| Change Δ = L − E | | −0.08 | −0.07 | +1.05 | 1.25 | −2.66 | −1.83 |
| T, 2020–2022 (descriptive) | 747 | −0.05 | −0.05 | 0.16 | 0.17 | 0.46 | 0.38 |
| All later sessions, 2020–Oct 2026 (descriptive) | 1,700 | 1.69 | 2.27 | 2.23 | 3.84 | 0.62 | 0.66 |
| All sessions, 2016–2026 | 2,677 | 2.22 | 4.45 | | | | |

The primary sample satisfies the no-weakening verdict: $\bar{R}_L$ = 3.05 bps with t = 2.98, and the change is −0.08 bps (t = −0.07). The 90% confidence interval for Δ runs from −1.90 to +1.74 bps; the one-sided 95% confidence bound on the decline is therefore 1.90 bps, or 61% of the 2016–2019 effect, and any larger decline is rejected at the 5% level. The quintile spread is consistent: 2.63 bps in L versus 2.55 in E.

The proxy sample (descriptive) addresses the main limitation of the primary sample, which cannot be rebuilt from outside. It gives 2.81 bps (t = 6.78) in 2016–2019, close to the primary estimate, and 3.86 bps (t = 5.26) in 2023–2026; the 2023–2026 estimate is thus larger, not smaller, in the sample anyone can rebuild from public data. The full index behaves differently: its 2016–2019 reversal of 3.40 bps falls to 0.74 bps (t = 0.55) in 2023–2026, a change of −2.66 bps (t = −1.83). Because dollar-volume weights favour the largest stocks, and those stocks carry more than half of the weight in the later period (Table 1), the full-index estimate is dominated by them. Because we had examined the excluded stocks before, their behaviour is not evidence in this design. The pre-registration nonetheless specified an appendix on the excluded stocks alone, reported here as registered and labelled as not evidence: their reversal was 3.76 bps (t = 5.27) in 2016–2019 and −0.09 bps (t = −0.05) in 2023–2026, a change of −3.86 bps (t = −1.97). The reversal among the 80 largest stocks by prior-month dollar volume—computable from public data—is 3.88 bps (t = 5.74) in 2016–2019, 1.19 bps (t = 0.76) in 2020–2022 and 0.33 bps (t = 0.18) in 2023–2026, a change of −3.55 bps (t = −1.81). This comparison is descriptive, and roughly 60% of those 80 stocks are in our excluded set. The paired daily difference between the proxy sample and these 80 stocks changes by +4.61 bps (t =

2.60), from −1.07 bps in 2016–2019 to +3.54 bps in 2023–2026.

Table 3 shows how far these statements depend on the window, the end point and the sample. A bound above 100% means that the confidence interval includes a complete disappearance of the reversal. On the one-sided test we use to bound the primary decline, three falls are significant: the primary sample over all sessions from 2020 ($p = 0.046$), the full index ($p = 0.034$) and the 80 largest stocks ($p = 0.035$). On the registered bar none is. We report both, because holding good and bad news to different bars would overstate persistence. Measured per unit of cross-sectional dispersion of LH, the primary reversal fell by about a third (3.13 per 22.8 bps of dispersion, against 3.05 per 33.3); "the same size" holds only in raw basis points.

*Table 3. Sensitivity to window, end point and sample (reversal, bps per day; descriptive; percentages use unrounded means).*

| cut | 2016–2019 | later | t, later | change | t, change | one-sided p | one-sided 95% bound on the decline |
|---|---|---|---|---|---|---|---|
| primary, 2023–2026, as registered | 3.13 | 3.05 | 2.98 | −2% | −0.07 | 0.472 | 61% |
| primary, 2023–2026, ending at 3:55 | 2.90 | 2.07 | 2.07 | −29% | −0.77 | 0.221 | 90% |
| primary, 2024–2026 | 3.13 | 1.80 | 1.37 | −43% | −0.97 | 0.166 | 115% |
| primary, 2020–2026 | 3.13 | 1.69 | 2.27 | −46% | −1.69 | 0.046 | 91% |
| proxy, 2023–2026 | 2.81 | 3.86 | 5.26 | +38% | 1.25 | | 12% |
| proxy, 2020–2026 | 2.81 | 2.23 | 3.84 | −20% | −0.80 | 0.211 | 62% |
| full index, 2023–2026 | 3.40 | 0.74 | 0.55 | −78% | −1.83 | 0.034 | 148% |
| 80 largest only, 2023–2026 | 3.88 | 0.33 | 0.18 | −92% | −1.81 | 0.035 | 175% |

### *4.3 Year by year*

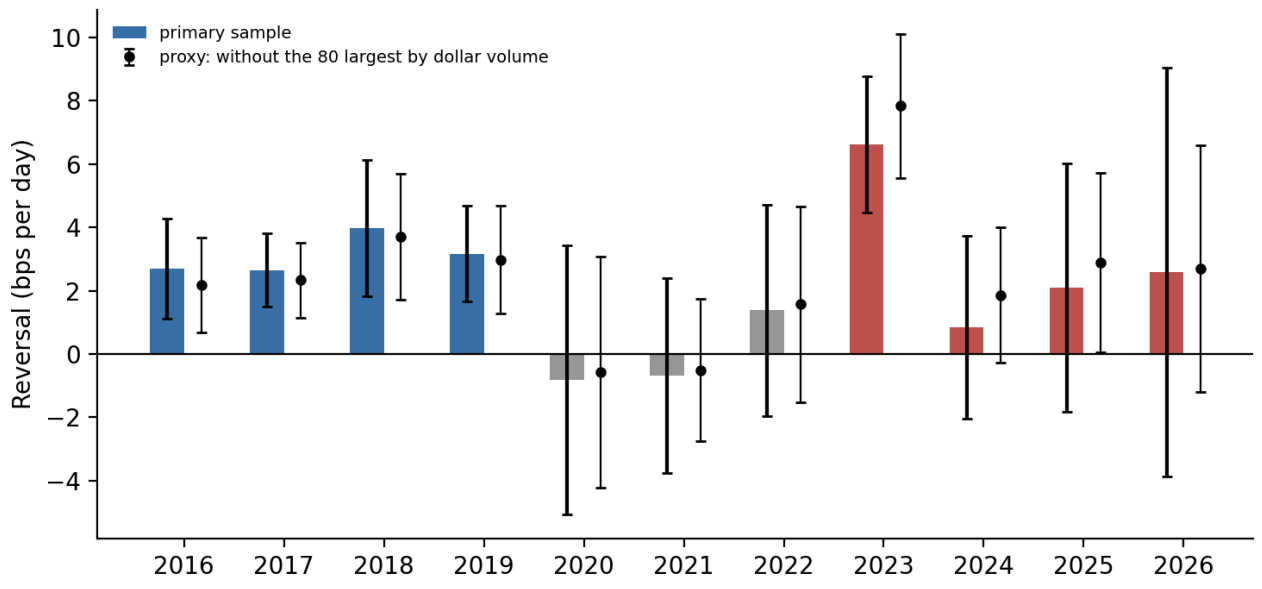

*Figure 1. End-of-day reversal by year, in basis points per day, with 95% Newey–West intervals. Bars show the primary sample: 2016–2019 (the reproduction test), 2020–2022 (descriptive) and 2023–2026 (the registered test period). Dots show the proxy sample that drops the 80 largest stocks by prior-month dollar volume. 2026 runs to October 30.*

*Table 4. Reversal by year, primary sample (bps per day; t in parentheses).*

| year | rank slope reversal | quintile spread |
|---|---|---|
| 2016 | 2.70 (3.34) | 2.22 (3.21) |
| 2017 | 2.65 (4.47) | 2.36 (4.42) |
| 2018 | 3.98 (3.64) | 3.03 (3.36) |
| 2019 | 3.17 (4.12) | 2.54 (4.00) |
| 2020 | −0.82 (−0.38) | −0.66 (−0.35) |
| 2021 | −0.68 (−0.44) | −0.23 (−0.18) |
| 2022 | 1.38 (0.81) | 1.34 (0.97) |
| 2023 | 6.62 (6.05) | 5.69 (6.26) |
| 2024 | 0.85 (0.58) | 0.76 (0.63) |
| 2025 | 2.10 (1.05) | 2.05 (1.21) |
| 2026 (to Oct 30) | 2.58 (0.78) | 1.90 (0.74) |

The reversal is positive and significant in each year from 2016 to 2019. In 2020 and 2021 it has the opposite sign, although neither year is significant, and the daily estimate is noisier in those years, with standard errors two to three times those of 2016–2019. The 2020–2022 mean is indistinguishable from zero, and the difference from 2016–2019 is 3.18 bps ($t = 2.79$; descriptive).

2023 is the strongest year in the sample and carries much of the 2023–2026 t-statistic, supplying 56% of the sum of the daily estimates in 2023–2026. Without it, the 2024–2026 mean is 1.80 bps ($t = 1.37$), the same sign and about 40% smaller; on its own, that sub-period would not meet the no-weakening bar of $t \geq 2$, and none of its years is significant. 2023 differs from 2024–2026 (gap $t = 2.82$), and 2024–2026 cannot be told apart from 2020–2022 (gap $t = 1.10$). The pre-registered stability check required at least six of the eight half-year windows in 2023–2026 to share the sign of the period mean; all eight do, with reversals from 0.75 bps (second half of 2024) to 8.22 bps (second half of 2023). In the proxy sample, all eight half-years are again positive. The two years after BDS's paper first circulated, 2025 and 2026, are positive in both samples but provide little power on their own.

### *4.4 Change-point tests*

A sup-Wald test for a single shift in the mean of the daily reversal over 2016–2026 (Andrews, 1993; 15% trimming, block-bootstrap p-value) gives p = 0.72, with the most likely date in April 2020 and a pre-registered 90% interval for the date from November 2017 to December 2024. The same test on linearly detrended data also gives p = 0.72. The pre-registered threshold for a break was p < 0.01 with a date inside 2020–2024; accordingly, we do not describe the series as having a break or a gradual change. The test does not reject a constant mean, but it is not suited to a dip that reverses: a single-break test compares the mean before and after one date, and a three-year interruption followed by a return to the earlier level leaves the two sides of most candidate dates similar. A simulation with the estimated period means and block-resampled residuals shows how little it can see: the sup-Wald statistic rejects at 5% in 19% of draws and at 1% in 6%, and the full registered rule almost never fires (below 1%). A test of equal means across the three periods (descriptive) rejects: $\chi^2(2) = 7.90$, p = 0.019 in the primary sample, p = 0.005 in the proxy and p = 0.026 on the full index. The period split was fixed in advance, but its boundaries match the dip in Figure 1, so this test only describes that figure.

### *4.5 The closing auction, a placebo and robustness*

Ending the window at the vendor's daily close instead of the last trade before 4:00 gives a reversal of 3.66 bps in E and 3.82 bps in L, a change of +0.16 bps (t = 0.14). The auction leg on its own, the return from the last trade before 4:00 to the vendor's daily close regressed on the same rank, also shows a reversal in every period: 0.53 bps in E (t = 15.2), 1.08 bps in T (t = 2.0) and 0.76 bps in L (t = 15.8). The auction leg was pre-registered as a reported quantity without a threshold and lies outside the robustness family below, so it affects no verdict. Rebuilt after the run from actual auction prints, the leg is 0.52 bps (t = 15.8), 0.40 bps (t = 6.9) and 0.76 bps (t = 15.8), and its increase from E to L is 0.24 bps (t = 4.09). The increase is therefore not an artefact of the vendor's close. The one large difference is in 2020–2022: on a few sessions in March 2020 the vendor's daily close was not the auction print, and the registered daily slope reached 85 to 256 bps, whereas on the auction print the largest daily slope in the whole sample is 15 bps. The auction leg has the same sign as the pre-auction reversal in every period, so the registered result is not an artefact of ending before the auction.

Table 5 (descriptive) splits the close into three legs on the same rank and weights, using the auction print. The auction leg grew in every sample, but the other legs disagree across samples. In the primary sample, the part earned before 3:55 fell from 2.90 to 2.07 bps and the last five minutes before the auction rose from 0.23 to 1.00 bps, neither significantly. In the proxy sample, the first leg did not fall and the last five minutes rose (+0.84 bps, t = 2.24). Among the 80 largest stocks, most of the 2016–2019 reversal sits in the last five minutes (2.90 of 3.88 bps), and that leg fell to 0.66 bps (change t = −2.23). BDS's pattern survives skipping the last five minutes over 1993–2019; among the largest stocks in 2016–2019, ours largely does not. We therefore do not claim that the reversal moved toward the close.

*Table 5. The close in three legs (reversal, bps per day; t in parentheses; descriptive). Each leg is a simple return, so the first two sum to the whole only up to a cross term of about 0.01 bps and rounding.*

| sample | leg | E, 2016–2019 | L, 2023–Oct 2026 | change | t(change) |
|---|---|---|---|---|---|
| primary | 3:30 to 3:55 | 2.90 (7.19) | 2.07 (2.07) | −0.83 | −0.77 |
| | 3:55 to last trade before 4:00 | 0.23 (1.42) | 1.00 (2.18) | +0.77 | 1.58 |
| | closing auction print | 0.52 (15.80) | 0.76 (15.78) | +0.24 | 4.09 |
| proxy | 3:30 to 3:55 | 2.69 (6.74) | 2.91 (4.27) | +0.22 | 0.28 |
| | 3:55 to last trade before 4:00 | 0.11 (0.74) | 0.95 (2.80) | +0.84 | 2.24 |
| | closing auction print | 0.43 (13.48) | 0.78 (22.47) | +0.36 | 7.60 |
| 80 largest | 3:30 to 3:55 | 0.98 (1.58) | −0.34 (−0.20) | −1.31 | −0.75 |
| | 3:55 to last trade before 4:00 | 2.90 (8.43) | 0.66 (0.70) | −2.24 | −2.23 |
| | closing auction print | 0.83 (12.11) | 1.05 (7.93) | +0.22 | 1.50 |
| full index | 3:30 to 3:55 | 1.80 (3.79) | 0.19 (0.15) | −1.61 | −1.22 |
| | 3:55 to last trade before 4:00 | 1.60 (6.80) | 0.55 (0.79) | −1.05 | −1.42 |
| | closing auction print | 0.63 (16.83) | 0.90 (9.76) | +0.26 | 2.64 |

As a placebo, the same design applied to the 2:00–2:30 p.m. return, sorted on the return to 1:30, gives a reversal of 1.39 bps in E (t = 2.98) and 0.38 bps in L (t = 0.45), a change of −1.02 (t = −1.07). In 2016–2019, a reversal less than half the closing size also appears in this earlier half-hour; in our sample the pattern is strongest at the close but not confined to it. The pre-registered placebo check asked whether the closing change exceeds the placebo change; because the closing change is close to zero, the check fails and the placebo cannot show that any change is specific to the close.

The robustness family consists of eight variants, Holm-corrected: the primary and full-index universes crossed with two end points (the last trade before 4:00 and the price at 3:55) and two statistics (rank slope and quintile spread), less the primary specification, plus the vendor's-close version of the primary. Table 6 reports them; none shows a significant change after correction, and the smallest Holm-adjusted p-value is 0.54. The full-index variants all show smaller reversals in L than in E, consistent with Section 4.2.

*Table 6. Pre-registered robustness variants (reversal, bps per day).*

| universe | end point | statistic | E | L | Δ | t(Δ) |
|---|---|---|---|---|---|---|
| primary | last trade before 4:00 | quintile | 2.55 | 2.63 | +0.08 | 0.09 |
| primary | 3:55 | rank slope | 2.90 | 2.07 | −0.83 | −0.77 |
| primary | 3:55 | quintile | 2.38 | 1.74 | −0.65 | −0.74 |
| full index | last trade before 4:00 | rank slope | 3.40 | 0.74 | −2.66 | −1.83 |
| full index | last trade before 4:00 | quintile | 2.79 | 1.17 | −1.62 | −1.44 |
| full index | 3:55 | rank slope | 1.80 | 0.19 | −1.61 | −1.22 |
| full index | 3:55 | quintile | 1.58 | 0.57 | −1.02 | −0.99 |
| primary | vendor's daily close | rank slope | 3.66 | 3.82 | +0.16 | 0.14 |

Table 7 (descriptive) shows what lies behind the full-index weakness. Winsorizing returns at 1% and 99% does not restore it (0.96 bps), so it is not about outlying returns. Equal weights (3.17 bps, t = 4.37), square-root weights (2.65 bps, t = 3.09) or dropping the five largest stocks each day (2.98 bps, t = 3.11) do restore it. Dollar-volume weights became far more concentrated: on the full index the top five stocks carry 13% of the weight in E and 27% in L, the largest stock 4% and 10%, and the effective number of stocks, $1/\sum w^2$, falls from 148 to 45. In the primary sample these figures are 6% and 9%, and 257 and 188, and in the proxy 3% and 3%, and 328 and 311. No single stock decides the primary or proxy result: leaving out any one stock, with the ranks kept fixed, gives 2.90 to 3.68 bps in the primary sample and 3.75 to 3.90 in the proxy for 2023–2026. On the full index the same range is 0.38 to 2.02 bps. We have no market-capitalization series, so we cannot report BDS's own weights. The choice of five stocks was made after looking. Dropping the largest 1, 3, 5, 10 or 20 stocks each day gives 2.72, 2.86, 2.98, 3.49 and 3.85 bps (t from 2.33 to 4.96), and the paired gap to the estimate with no stock dropped is significant at each cut (t from 2.64 to 3.22); in 2016–2019 the same gaps are 0.07 to 0.25 bps. The largest stock alone, a different one on different days, accounts for 1.97 of the 2.24 bps gap to the drop-five estimate. The 2020–2022 dip does not depend on the weights: that period's reversal stays within 0.2 bps of zero in the primary and proxy samples under every row of Table 7.

*Table 7. Weights and outliers (reversal in 2016–2019 / 2020–2022 / 2023–2026, bps per day; in parentheses the t for 2023–2026, then the t of the change from 2016–2019 to 2023–2026; descriptive).*

| | primary | proxy | 80 largest | full index |
|---|---|---|---|---|
| dollar volume (as registered) | 3.13 / −0.05 / 3.05 (2.98; −0.07) | 2.81 / 0.16 / 3.86 (5.26; 1.25) | 3.88 / 1.19 / 0.33 (0.18; −1.81) | 3.40 / 0.46 / 0.74 (0.55; −1.83) |
| square root of dollar volume | 2.82 / 0.05 / 3.40 (4.43; 0.67) | 2.65 / 0.16 / 3.64 (5.10; 1.20) | 4.19 / 1.68 / 1.98 (1.42; −1.47) | 3.17 / 0.59 / 2.65 (3.09; −0.54) |
| equal | 2.58 / 0.09 / 3.27 (4.58; 0.85) | 2.51 / 0.16 / 3.35 (4.78; 1.04) | 4.45 / 1.98 / 2.97 (2.49; −1.14) | 2.85 / 0.43 / 3.17 (4.37; 0.39) |
| LH winsorized at 1%, 99% | 3.00 / −0.04 / 3.13 (3.62; 0.14) | 2.69 / 0.12 / 3.60 (5.05; 1.11) | 3.76 / 1.52 / 0.69 (0.40; −1.68) | 3.28 / 0.73 / 0.96 (0.79; −1.77) |
| dropping the 5 largest each day | 3.04 / −0.02 / 3.80 (4.89; 0.86) | 2.68 / 0.12 / 4.01 (5.50; 1.59) | 4.54 / 2.60 / 2.84 (2.07; −1.15) | 3.65 / 1.42 / 2.98 (3.11; −0.63) |

### *4.6 The next morning*

BDS find that the reversal is undone the following day: from 3:30 p.m. to the next day's open, their coefficient on ROD3 moves from −0.69 (t = −5.78) to −0.12 (t = −0.47), a give-back of about 83%. Our pre-registered price-pressure check measured the next-morning return from the last trade before 4:00 to the next day's open, and defined κ as the mean next-morning slope divided by the mean closing slope; it required κ to lie in [−1.2, −0.3] with a combined slope close to zero. The check fails on the primary sample in both periods. In 2016–2019, κ = 0.62 (90% interval 0.15 to 1.45): the next morning continues the closing reversal rather than giving it back, and the combined reversal is 5.07 bps against 3.13 for the close alone. This clashes with BDS, and the auction does not explain it, because their window includes the auction as well. In 2023–2026 the point estimate, κ = −1.49, is a give-back larger than the closing move, but the interval (−5.72 to 1.09) is too wide to read.

A descriptive comparison points to the sample. On the full index in 2016–2019, in a form close to BDS's, our coefficient moves from −0.83 (t = −7.33) to −0.07 (t = −0.22) by the next open, close to BDS's own figures, and κ on the rank slope is −0.91, inside the registered range. The form differs from theirs: ours averages daily regressions on the raw return, winsorized at 1% and 99%, with dollar-volume weights and Newey–West errors, whereas BDS estimate a panel with date and firm effects, capitalization weights and errors clustered by date and firm. In the primary and proxy samples in 2016–2019 the coefficient moves the other way, from −0.77 to −1.40 and from −0.65 to −1.37, so in that period the primary sample does not look like the transitory price pressure they describe. The full-index match may be a mix and not a give-back in any one group: its combined reversal of 0.30 bps in 2016–2019 (Table 8) lies between 5.51 bps in the proxy sample and −5.10 bps among the 80 largest stocks. We offer no mechanism for this difference, and the samples are not directly comparable: BDS cover 27 years and we cover four, and our window ends at the vendor's open rather than at the opening-auction print, which differ by 3 to 9 bps on average, although ending at the print changes little (Table 8).

In 2023–2026 the full-index next morning is much larger than the closing move: the combined reversal is −13.7 bps (t = −2.88), so intraday winners beat intraday losers from 3:30 p.m. to the next open. This is the largest number in the paper, and Table 8 checks it. Split and dividend adjustments do not explain it. The window crosses sessions on prices adjusted for splits and dividends: on the 98 split-size nights the median absolute return is 0.9%, where raw prices would give 50%, and on the 22,522 dividend-size nights the adjustment has a median size of 41 bps while the adjusted return stays in line with other nights. Removing all split-size nights leaves −13.72 bps, removing every night with an adjustment gives −13.89 bps, and removing every next-morning move above 20% gives −13.91 bps (Appendix C). The real opening print gives −13.70 bps. It does not rest on a few days: the 10 largest days by size sum to +551 bps against a total of −13,057 bps over 952 days, so without them it is −14.45 bps, and the median day is −11.67 bps. It does rest on the weights of a few stocks. It falls to −9.64 bps when winsorized, −0.93 with square-root weights and −1.41 without the five largest stocks each day, and it is +3.75 with equal weights. Leaving out one stock can move it to −6.23. Among the 80 largest stocks it is −20.79 bps (t = −3.22), and in the proxy sample it is +6.15 bps (t = 2.66). The primary sample lies between them at −1.37 bps (t = −0.32); with κ = −1.49 its point estimate is a give-back, but the interval does not separate it from no give-back, and its 10 largest days supply nearly all of it (−0.07 bps without them). We do not test why it occurs.

*Table 8. From 3:30 p.m. to the next open (reversal, bps per day, compounded; t in parentheses; negative = intraday winners beat losers; descriptive). The last five columns are 2023–2026. The registered check in the text adds the closing and next-morning returns (5.07 bps); this table compounds them (5.08 bps). κ is a ratio of slopes and is not a figure in this table.*

| | 2016–2019 | 2020–2022 | 2023–2026 | real opening print | LH to open winsorized | square-root weights | equal weights | without 5 largest |
|---|---|---|---|---|---|---|---|---|
| full index | 0.30 (0.19) | −12.03 (−2.16) | −13.72 (−2.88) | −13.70 (−2.87) | −9.64 (−2.42) | −0.93 (−0.31) | 3.75 (1.60) | −1.41 (−0.43) |
| 80 largest | −5.10 (−2.30) | −13.04 (−2.04) | −20.79 (−3.22) | −20.63 (−3.20) | −17.69 (−3.01) | −10.24 (−2.07) | −3.97 (−0.92) | −5.13 (−1.11) |
| primary | 5.08 (3.90) | −10.36 (−2.01) | −1.37 (−0.32) | −1.19 (−0.28) | 1.52 (0.43) | 3.30 (1.18) | 4.88 (2.14) | 4.67 (1.69) |
| proxy | 5.51 (4.24) | −4.66 (−1.11) | 6.15 (2.66) | 5.82 (2.45) | 6.53 (3.01) | 5.95 (2.70) | 5.69 (2.70) | 5.99 (2.64) |

## 5. Mechanism tests

### *5.1 The pre-registered family*

The four mechanism tests form one Holm (1979) family at α = 0.05, each two-sided, and a mechanism is described as supported only if its test passes. The first two split ROD3 into a market component and an idiosyncratic component and estimate both in one regression,

$$\mathrm{LH}_{i,t} = a_t + b_t^m\, q\big(\beta_{i,t}\,\mathrm{ROD3}_t^{SPY}\big) + b_t^e\, q\big(\mathrm{ROD3}_{i,t} - \beta_{i,t}\,\mathrm{ROD3}_t^{SPY}\big) + e_{i,t}, \tag{4}$$

where q(·) is the scaled cross-sectional rank, $\mathrm{ROD3}_t^{SPY}$ is the same window for the SPDR S&P 500 ETF, and $\beta_{i,t}$ is the slope of the stock's daily return on the ETF's daily return over the previous 252 sessions (at least 200), using data up to day t − 1. Because the market's move is the same for every stock on a given day, the market component is a rank on β: high-β stocks sit at the top on up days and at the bottom on down days. Because the β estimate needs about a year of history, these tests start in late 2016. The third test is a difference-in-differences regression with Newey–West errors on sessions from 2017 onward. It compares the market-component reversal on Tuesdays and Thursdays, the weekdays that gained same-day S&P 500 index option expiries in April and May 2022, with the reversal on the other weekdays, before and after May 2022. The fourth is a stacked difference-in-differences around each stock's first single-stock leveraged-ETF launch, described in Section 5.2.

*Table 9. Mechanism tests (reversal, bps per day; the third and fourth rows report effects; period E starts in late 2016 because β needs a year of history).*

| test | E | L | change or effect | t | Holm |
|---|---|---|---|---|---|
| market component, $-b^m$ | −0.34 | 2.39 | +2.73 | 1.78 | not passed |
| idiosyncratic component, $-b^e$ | 3.39 | 2.51 | −0.87 | −0.92 | not passed |
| Tue/Thu after May 2022, market component | | | −1.45 | −0.50 | not passed |
| first single-stock leveraged-ETF launch | | | not judged (12 treated stocks, 20 required) | | not passed |

The market component moved from slight continuation in 2016–2019 to a reversal of 2.39 bps in 2023–2026, a change of 2.73 bps (t = 1.78, p = 0.075). This is the direction predicted by a dampening effect of index option hedging, but it does not clear the threshold. The idiosyncratic component remained a reversal in both periods with no significant change. The weekday test shows no effect and is weak by design: same-day SPX expiries existed on every weekday after May 2022, and SPY and QQQ options added Tuesday and Thursday expiries in November 2022 (Cboe, 2022); the comparison therefore measures only how recently each weekday gained same-day index options, and its power is low. The leveraged-ETF test required at least 20 treated stocks inside the sample, and there are 12, because most single-stock leveraged ETFs are written either on the largest stocks, which our design excludes, or on stocks that joined the index too late to provide a 250-session window. Under the pre-registered rule, that test is exploratory and counts as not passed.

### *5.2 An exploratory analysis of leveraged-ETF launches*

We built the launch table before the confirmatory run from issuer documents, including product guides, fact sheets, annual reports and launch releases. A first version relied on an aggregator website, and a spot check showed that the site used one placeholder date for many funds of a single issuer, so every row was re-dated from issuer documents and nine dates changed before the run (Appendix B). For each treated stock, the stock's closing slope over a window is the ordinary least-squares slope of its own LH on its daily cross-sectional rank q, estimated over its sessions in that window. The test compares this slope in the 250 sessions after the stock's first leveraged-ETF launch with the 250 sessions before, and subtracts the same change for three untreated stocks of similar dollar volume. The registered script reports the pooled estimate over all 34 index members with a first launch from July 2022 and a complete 250-session post-launch window; the registered launch cutoff was December 2025, and the latest such launch is in October 2025. The reversal falls by 3.5 bps (t = −1.3), the predicted direction. Under the pre-registered rule this estimate is exploratory, because fewer than 20 of the 34 lie in the primary sample.

Our primary sample contains 12 of the 34 treated stocks. For those 12, the reversal falls by 13.1 bps, and it falls for every one of them; for the 22 treated stocks outside our sample, it rises by 1.7 bps. Because this split was chosen after the confirmatory run, we treat it as a hypothesis for future testing rather than as evidence. It has three further weaknesses. Issuers launch leveraged ETFs on stocks that are already attracting demand, and a change in that demand could move the closing slope by itself. Ten of the 12 launches fall in 2025, so the post-launch windows overlap heavily and the stocks are not independent observations. We report no t-statistic for this split: the spread of the 12 estimates looks smaller than sampling noise alone would produce, which points to a common shock or to the shared control stocks. And twelve stocks are too few to rule out chance.

About 35 index members received their first leveraged ETF after December 2025, and their 250-session post-launch windows are not yet complete. Five launches in November and December 2025 fall between the two groups: their windows were incomplete on October 30, 2026, and the registered test starts in January 2026. We registered a test on the 35 later stocks on November 5, 2026, and deposited it on November 6, 2026, before computing any stock-level result for these names, and the registration is deposited in the public archive: it uses the same estimator on the stocks of the primary sample, predicts a smaller reversal after launch, and requires $t \geq 3$ on at least 15 stocks. It will run once, after the last post-launch window closes, around the end of 2027. From membership alone, 32 of the 35 stocks are in the primary sample, which should satisfy the 15-stock requirement, although their launch dates must still be confirmed from issuer documents before the test runs. We do not name the primary-sample stocks, because that would identify excluded ones; a dated addendum in the archive therefore adds the same estimate on all treated stocks, which anyone can rebuild, and a descriptive row for the five stocks in between.

## 6. Limitations

The main result is measured before the closing auction; the auction leg has the same sign in every period, but we do not combine the two legs. It covers S&P 500 members and says nothing about small stocks, for which BDS find a stronger effect in equal-weighted portfolios, and with dollar-volume weights it says nothing positive about the very largest stocks. The primary sample cannot be rebuilt exactly from outside because the exclusion list is not released, and because the list reflects recent trading it looks ahead in 2016–2019. The earlier work that led to the list is unpublished. The proxy sample is our answer to these points, but it was chosen after the confirmatory run, it shares most of its stocks with the primary sample, and it is not part of the registered evidence.

The result depends on the window, the end point and the sample (Table 3). The daily reversal was noisier in 2023–2026 than in 2016–2019, and the standard error of the change, 1.11 bps, is twice the 0.56 bps projected at the reproduction test, which combined its standard error of 0.42 with a projected 0.37 (the 2016–2019 daily standard deviation over 960 sessions). The verdicts are asymmetric: "weakened" needed $t \leq -3$ on the change, while the no-weakening verdict needed only $t \geq 2$ on the level. A true decline of 50% would reach the "weakened" bar with a probability of only about 6%, and it would produce the no-weakening verdict about 32% of the time (Table A2), so the design could confirm persistence far more easily than it could detect a decline. The change-point test has almost no power against a dip that reverses (Section 4.4). The placebo and next-morning checks both failed on the primary sample, and neither supports a closing-specific price-pressure interpretation there. The weekday option test has low power by construction, and the leveraged-ETF test could not be judged. The prices come from a single vendor and were not cross-checked against a second source; the auction-print comparison of Section 3.2 uses the same vendor. The outcome uses trade prices, not quotes: if intraday losers end on buyer-initiated trades, the last trade sits nearer the ask and LH for losers is overstated by up to half a spread. BDS report that their result survives in quote midpoints; we have not measured spreads or midpoints for our sample, although the auction leg, in which losers rise further into the print, does not fit this artefact. Our weights are dollar volume, not market capitalization, and the full-index result changes sharply with the weights (Table 7). The pre-registration timestamps come from a private version history. Finally, the pattern amounts to a few basis points a day before costs, and we make no claim about returns net of costs.

## 7. Conclusion

The end-of-day reversal of Baltussen, Da and Soebhag was present in 2023–2026 in S&P 500 stocks outside the heavily traded names we excluded. Intraday losers beat intraday winners over the last half hour by about 3 bps a day, the same size in raw basis points as in 2016–2019 and about a third smaller per unit of dispersion, and a proxy sample that anyone can rebuild gives a larger estimate. The persistence was not uninterrupted: the reversal was indistinguishable from zero in 2020–2022, it was strongest in 2023, and over all sessions from 2020 the one-sided 95% bound on the decline is 91%. On the full index with dollar-volume weights, the 2023–2026 reversal is small and insignificant, but it comes back with equal weights or without the five largest stocks each day, which suggests that the very largest stocks behave differently, or that a few of them have dominated the weights, and not that the reversal has left the index. Those comparisons were made on data we had already seen and therefore cannot be tested on 2023–2026 again; only sessions after October 2026 can test them, and such a test could use the proxy sample as its primary sample because anyone can rebuild it. No pre-registered mechanism test explains these movements. In 2016–2019, BDS's next-morning give-back appears on the full index and does not appear on the primary sample, although the full-index figure mixes opposite behaviour in the largest stocks and the rest. In 2023–2026 the primary estimate is too noisy to read, and the largest stocks continue overnight by far more than the closing move, which leaves the price-pressure account open. Two further questions remain open: why the reversal was absent in 2020–2022, a period that includes the 2020 crash, near-zero interest rates and a surge in retail trading, none of which we test, and whether leveraged-ETF launches weaken it stock by stock. The second can be answered with launches whose post-launch windows are not yet complete and whose stock-level results have not been computed.

## Appendix A. Pre-registered decision rules

The rules were written in terms of the daily slope $b_t$, with $\Delta_b = \bar{b}_L - \bar{b}_E$. Since $R_t = -b_t$, each rule below is given in the original form and in the reversal form used in the text.

*Table A1. Decision rules.*

| rule | as registered (slope) | in reversal terms |
|---|---|---|
| reproduction test | $\bar{b}_E < 0$, $t \leq -2$ | $\bar{R}_E > 0$, $t \geq 2$ |
| weakened | $\Delta_b > 0$, $t \geq 3$ | $\Delta < 0$, $t \leq -3$ |
| changed sides | weakened, $\bar{b}_L > 0$ at $t \geq 3$, vendor's daily close agrees | weakened, $\bar{R}_L < 0$ at $t \leq -3$, vendor's daily close agrees |
| no weakening | $\Delta_b \leq 0$, or $\bar{b}_L < 0$ with $t \leq -2$ | $\Delta \geq 0$, or $\bar{R}_L > 0$ with $t \geq 2$ |
| order of precedence | changed sides first; then, if weakened and no weakening both hold, weakened but still a reversal; then weakened alone, no weakening alone, not proven | same |
| break | sup-Wald $p < 0.01$, date in 2020–2024, detrended test also $p < 0.01$ | same |

| rule | as registered (slope) | in reversal terms |
|---|---|---|
| mechanisms | one Holm family of four two-sided tests at 0.05; leveraged-ETF test judged only with ≥ 20 treated stocks in the sample | same |
| stability | at least 6 of 8 half-years in L with the sign of the L mean; a random-halves check applies only if the change is significant | same |
| robustness | eight variants, Holm-corrected; they cannot rescue a failed primary result | same |

*Table A2. Verdicts simulated at the realised standard errors (share of 400,000 draws; added after the run).*

| true decline | "weakened" (with or without a remaining reversal) | "no weakening" | "not proven" |
|---|---|---|---|
| 0% | 0% | 85% | 15% |
| 25% | 1% | 62% | 37% |
| 50% | 6% | 32% | 62% |
| 75% | 19% | 11% | 70% |
| 100% | 43% | 2% | 55% |

## Appendix B. Changes made before any primary-sample outcome was seen

The adversarial audit found that a linear drift with no break passed the original break test in 45% of simulated draws, so a detrended version was added and a break now requires both tests. It also found that a decline leaving a significant reversal satisfied both the "weakened" verdict and the no-weakening verdict, so an explicit order of precedence and the verdict "weakened but still a reversal" were added. Every claim was then worded as "before the closing auction", the auction leg was added as a reported quantity, and "changed sides" was made to require agreement from the version ending at the vendor's daily close. The leveraged-ETF launch table was first built from an aggregator; a spot check against issuer documents found a placeholder date shared by many funds of one issuer, all rows were re-dated from issuer documents, nine changed, and the table was re-hashed before the confirmatory run. Finally, a data panel built during quality checks contained later-period returns; no statistic relating ROD3 to LH was computed on it, and it was moved out of the working directory before the reproduction test and disclosed.

## Appendix C. Data checks

The weekdays without a session are exactly the exchange holidays plus two national days of mourning. Six sessions have vendor gaps and are dropped. Overnight returns use adjusted daily bars and intraday ratios use raw bars from the same session, so splits and dividends do not enter the intraday returns. The vendor's daily open matches the open of the first 30-minute bar within 1 bp on 90% of stock-days. Of the 46 stock-days with a move larger than 40%, most are genuine; the rest are spin-offs or ticker splices, and the overnight screen removes them. That screen applies to the overnight return in the sorting variable, not to the next-morning window. In that window 14 stock-days move more than 40%. Four nights, all before 2021 and outside the 80 largest stocks, look like spin-offs where the vendor applied a distribution ratio as a split, and one night of −81% in 2026 has no adjustment and may be a spin-off the vendor left unadjusted. We keep them, as registered, and report what removing them does. Removing every night with an adjustment moves the primary 2016–2019 combined reversal from 5.08 to 5.06 bps, κ from 0.62 to 0.61, the primary 2023–2026 κ from −1.49 to −1.51, and the full-index κ from −0.91 to −0.90; the full-index BDS-form coefficients move from −0.83 and −0.07 to −0.83 and −0.10. Removing every next-morning move above 20% gives 5.13 bps and κ of 0.64, −1.50 and −0.82.

## Appendix D. Analyses added after the confirmatory run

None of the following was pre-registered, and none enters a verdict. Each was computed after the confirmatory run, with the registered filters and estimator. The first set came from a separate script; the later sets import the registered estimator functions and do not run the registered script again. The proxy sample that drops the 80 largest members by prior-month dollar volume (Sections 3.1 and 4.2), and the reversal among those 80 stocks alone (Section 4.2), were added because the excluded list is not released. The full-index t-statistics for 2016–2019 and 2020–2022 and the 2020–2026 row (Table 2), the sample description (Table 1), the weight share of the excluded stocks and the overlap between the excluded list and the 80 largest stocks were added for description. The comparison of 2016–2019 with 2020–2022 (Section 4.3) is descriptive. The following were added in response to outside reviews: the 2024–2026 mean without 2023, 2023's share of the 2023–2026 sum and the gap tests (Section 4.3); the sensitivity table (Table 3); the paired difference between the proxy sample and the 80 largest stocks (Section 4.2); the three-period test and the power simulation (Section 4.4); the t-statistics of the placebo levels and the rebuilt auction leg (Section 4.5); the comparison of the vendor's close and open with auction prints (Section 3.2); the sort rebased on the auction print (Section 3.3); the legs of the close (Table 5); the weight and outlier variants, concentration and leave-one-out figures (Table 7); the smallest Holm-adjusted p-value of the robustness family (Section 4.5); the cuts dropping 1 to 20 stocks and the 2020–2022 column of Table 7 (Section 4.5); the full-index and proxy next-morning figures in a form close to BDS's and the next-morning rerun (Section 4.6, Table 8); the simulated verdicts (Table A2) and the power figure of Section 6. The count of 32 primary-sample stocks among the 35 with a first launch after December 2025 (Section 5.2) comes from membership alone. The split of the leveraged-ETF test by sample (Section 5.2) was chosen after the confirmatory run; its one test on later launches is registered and has not been run.

**Use of AI tools.** Claude (Anthropic) proposed and drafted the design, the pre-registration and the decision rules, which the authors adopted. An AI agent ran the adversarial audit. Claude also wrote and audited the code, ran the pre-registered analyses, and drafted and revised the text, and other AI models reviewed drafts. The authors reviewed, edited and verified the output, and take full responsibility for the content.

**Disclosure.** The authors do not trade a strategy based on the end-of-day reversal and received no external funding for this work.